\documentclass[12pt, review]{elsarticle}

\usepackage{hyperref}
\usepackage{amsmath, amsthm, amssymb, calrsfs, wasysym, verbatim, bbm, color, graphics, geometry, siunitx, setspace, float, caption}

\newfloat{Video}{H}{lop}
\floatname{Video}{Video}

\journal{Scripta Materialia}

\begin{document}

\begin{frontmatter}

%% Title, authors and addresses

%% use the tnoteref command within \title for footnotes;
%% use the tnotetext command for theassociated footnote;
%% use the fnref command within \author or \address for footnotes;
%% use the fntext command for theassociated footnote;
%% use the corref command within \author for corresponding author footnotes;
%% use the cortext command for theassociated footnote;
%% use the ead command for the email address,
%% and the form \ead[url] for the home page:
%% \title{Title\tnoteref{label1}}
%% \tnotetext[label1]{}
%% \author{Name\corref{cor1}\fnref{label2}}
%% \ead{email address}
%% \ead[url]{home page}
%% \fntext[label2]{}
%% \cortext[cor1]{}
%% \affiliation{organization={},
%%             addressline={},
%%             city={},
%%             postcode={},
%%             state={},
%%             country={}}
%% \fntext[label3]{}

\title{Three\mbox{-}dimensional morphology of an ultrafine Al-Si eutectic produced \textit{via} laser rapid solidification}

\author[ummatphysaffil]{Xinyi Zhou \fnref{equalcontribution}}

\author[ummseaffil]{Paul Chao \fnref{equalcontribution}}

\fntext[equalcontribution]{These authors contributed equally: Xinyi Zhou and Paul Chao}

\author[ummseaffil]{Luke Sloan}

\author[ummseaffil]{Huai\mbox{-}Hsun Lien}

\author[mc2affil]{Allen H. Hunter}

\author[ummseaffil]{Amit Misra}

\author[ummseaffil,umcheaffil]{and Ashwin J. Shahani\corref{mycorrespondingauthor}}
\cortext[mycorrespondingauthor]{Corresponding author. Tel.: +1 (734) 764-5648; 2300 Hayward St, Ann Arbor, MI, 48109-2117 USA}
\ead{shahani@umich.edu}

\address[ummatphysaffil]{Engineering Physics Program, University of Michigan, Ann Arbor, MI 48109, USA}
\address[ummseaffil]{Department of Materials Science and Engineering, University of Michigan, Ann Arbor, MI 48109, USA}
\address[mc2affil]{Michigan Center for Materials Characterization, University of Michigan, Ann Arbor, MI 48109, USA}
\address[umcheaffil]{Department of Chemical Engineering, University of Michigan, Ann Arbor, MI 48109, USA}

\begin{abstract}
%% Text of abstract
Al\mbox{-}Si alloys processed by laser rapid solidification yield eutectic microstructures with ultrafine and interconnected fibers. Such fibrous structures have long been thought to bear resemblance to those formed in impurity\mbox{-}doped alloys upon conventional casting. Here, we show that any similarity is purely superficial. By harnessing high\mbox{-}throughput  characterization  and computer vision techniques, we perform a three\mbox{-}dimensional analysis of the branching behavior of the ultrafine eutectic and compare it against an impurity\mbox{-}modified eutectic as well as a random fractal (as a benchmark).  Differences in the branching statistics point to different microstructural origins of the impurity- and quench\mbox{-}modified eutectic. Our quantitative approach is not limited to the data presented here but can be used to extract abstract information from other volumetric datasets, without customization. 

%nanometer scal
%Characterizing the refined morphology of Al-Si eutectics under different modifications enables us to acquire desired mechanical properties of alloys in applications. In this paper, we characterize the ultra-fine fibrous Al-Si eutectic obtained in the center of the melt pool under laser surface remelting using 3D FIB tomography. We compare the morphological characteristics of these quench-modified structures to Si fibers observed with impurity modification and a random fractal simulated through Diffusion Limited Aggregation (DLA). The Si fibers formed under quenching showed a fundamentally different tree morphology towards impurity-modified ones regarding both a finer length scale and a isotropic branching behavior.
%The characterization and subsequent detailed analysis adds to our understanding of the mechanisms behind modifications during solidification processing.

\end{abstract}

%%Graphical abstract
%\begin{graphicalabstract}
%\includegraphics{grabs}
%\end{graphicalabstract}

%%Research highlights
%\begin{highlights}
%\item Research highlight 1
%\item Research highlight 2
%\end{highlights}

\begin{keyword}
%% keywords here, in the form: keyword \sep keyword
Rapid solidification \sep Laser surface remelting \sep 3D microstructure \sep eutectic%eyword two
\end{keyword}

\end{frontmatter}

%% main text
%\section{Introduction}
%\label{intro}

% @PC
% Motivation:  Importance of studying the refining process and resulted fine structures of Al-Si alloy system in 3D

Al\mbox{-}Si eutectic alloys are widely used as structural materials in automotive and aerospace applications due to their high strength\mbox{-}to\mbox{-}weight ratio and excellent castability \cite{hellawell_growth_1970}. Their mechanical properties such as tensile strength and ductility can be further improved by avoiding the formation of coarse, brittle Si flakes during solidification \cite{tiwary_based_2015, lien_plastic_2022,ferrarini2004microstructure,suryawanshi2016simultaneous}. This is accomplished \textit{via} \textit{impurity-modification}, in which the introduction of a few hundred ppm Na or Sr leads to a refinement of the Si into a fibrous structure on the order of micrometers \cite{yilmaz_microstrucutre_1989,moniri2019chemical}.  On the other hand, rapid solidification alone gives rise to a microstructural refinement \cite{roehling2017rapid, zhao2014ultra, hearn2018microstructure, bendijk1980characterization, gremaud1996development}: for example, laser surface remelting  creates rapid cooling conditions (as high as 10$^{10}$~K/s~\cite{von1984melt}) to refine the microscale flakes to nanoscale fibers \cite{pierantoni_coupled_1992, lien_microstructure_2020,abboud2020developing,dinda2012evolution}. Such a \textit{quench\mbox{-}modification} is promoted above a critical solidification front velocity~\cite{steent1972structure,hosch_analysis_2009}. In this context, the term ``modification" indicates a dramatic alteration in the morphology of Si that is produced by eutectic solidification.

It has generally been assumed by some authors \cite{steent1972structure,day1968microstructure,gruzleski1990treatment,tenekedjiev1990hypereutectic,nogita2004modification} that the two routes of modification lead to the same end\mbox{-}product (namely, fibers of Si in a matrix of Al). Yet this assumption has never been formally tested. Two\mbox{-}dimensional (2D) images of fibrous Si (see \textbf{Fig.~\ref{S0_compare}} for an example) may present similar features (\textit{e.g.}, aspect ratios~\cite{hosch_analysis_2009}), but the morphology and topology of the Si phase may be distinct in three\mbox{-}dimensions (3D). Importantly, Requena and coworkers~\cite{requena20093d,requena2011effect} have found that the load\mbox{-}carrying capacity of Al\mbox{-}Si alloys depends on the internal 3D architecture, \textit{i.e.}, the connectivity of the eutectic Si phase. Only recently is it possible to obtain a detailed characterization and quantitative, comparative analysis of these structures at the \textit{nanometric} level, due to advances in high resolution and high throughput 3D imaging and development of novel data\mbox{-}driven microstructure analysis tools. Computer vision provides an opportunity to construct rich 3D microstructural fingerprints~\cite{gaiselmann_competitive_2013}, which can be analyzed and compared for the two mentioned processing routes, and without significant assumption or human intervention.

%However, due to the limitation of existing processing methods, people are not able to extract morphological details from the microstructures.

% Elaborate
%during solidification for off\mbox{-}eutectic composition alloys 

Here, we leverage the above capabilities to quantify the 3D morphology of ultra\mbox{-}fine Si formed by laser quench-modification \cite{lien_microstructure_2020} and compare to ($i$) Si formed by  impurity-modification in directional solidification (previously characterized in \cite{gaiselmann_competitive_2013}) and to ($ii$) a random fractal simulated \textit{via} diffusion limited aggregation (DLA). \textcolor{black}{The laser remelted alloy was of  composition Al-20wt\%Si, cast from Al (99.99\%) and Si (99.99\%), see Ref.~\cite{lien_microstructure_2020} for additional details. The reason we selected a hyper-eutectic alloy for our quench experiment was to ensure that we remain within the (skewed) coupled zone of Al-Si~\cite{pierantoni_coupled_1992}, at the relatively high velocities associated with laser rapid solidification.} In case ($i$), the \textcolor{black}{directionally solidified} Al-\textcolor{black}{7wt\%}Si alloy contains 150 ppm of Sr \cite{gaiselmann_competitive_2013} \textcolor{black}{(\textit{i.e.}, above the concentration threshold for Sr modification~\cite{eiken2015impact}) and indeed it} shows typical impurity-modified morphological characteristics and the basic parameters listed in \textbf{Table~\ref{table}}. Our analysis of the branching behavior\footnote{\textcolor{black}{The branched structures presented herein resemble at first glance a dendritic morphology, but note they are much smaller than the primary Si dendrites observed in, \textit{e.g.},  Ref.~\cite{toropova2022microstructure}.}} offers insight into their morphological differences and  formation mechanisms.  % More broadly, our approach is not limited to the data presented here but can be extended to other volumetric datasets without customization. 

%\section{Methods} \label{methods}

\begin{figure}[h]
    \centering
    \includegraphics[scale=.56]{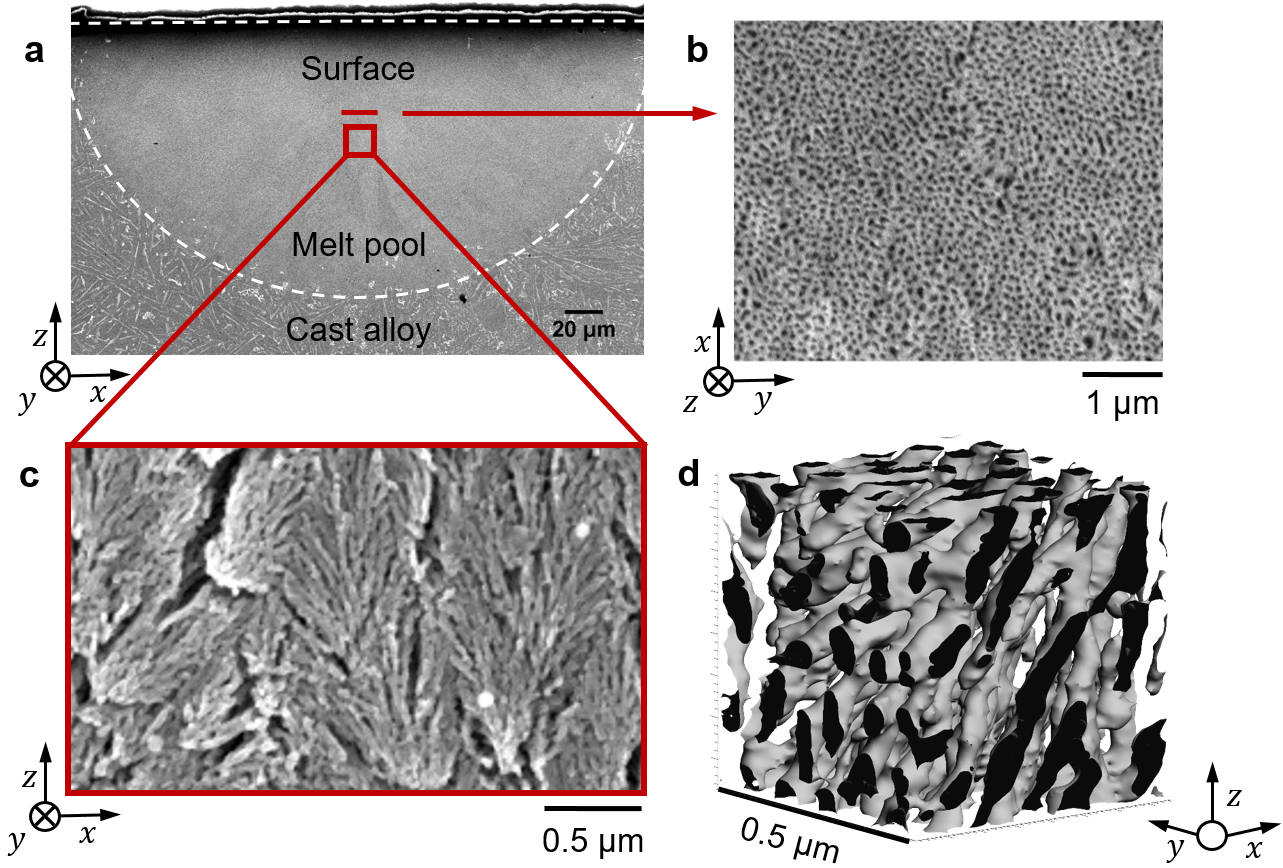}
    \caption{\textit{LSR produces an ultra\mbox{-}fine Al\mbox{-}Si eutectic microstructure} imaged by SEM\mbox{-}FIB tomography: (a) melt pool boundary (demarcated by white dashed line) and the location we selected for FIB tomography (red box). The laser moves along the $y$ direction at a speed of 30~mm/s. (b) SEM image of a FIB cross\mbox{-}section from the fibrous eutectic microstructure at the center of the melt pool. Dark spots are the cross\mbox{-}sections of Si fibers while the lighter regions correspond to the Al matrix. (c) An etched region that illustrates the coral\mbox{-}like structure within the melt pool, reproduced with permission from \cite{lien_microstructure_2020}.  (d) 3D reconstruction that displays the complex and bicontinuous morphology of eutectic Si structure. 
}
    \label{F1_FIB}
\end{figure}

%\subsection{3D characterization}
% @PC
To characterize the connectivity of fine fibrous Si, we utilize focused ion beam (FIB) tomography. Through serial\mbox{-}sectioning, we can reconstruct the 3D microstructure  \textit{via} scanning electron microscopy (SEM) \cite{uchic_threedimensional_2007}. This method is advantageous since it can accommodate volumes larger than~\SI{1000}{\micro\meter^3} with voxels approaching tens of nm compared to X\mbox{-}ray nanotomography, in which the sample dimension is limited to $\sim$\SI{50}{\micro\meter}~\cite{uchic_threedimensional_2007, shahani2020characterization}. Also, the similar x-ray attenuation coefficients of Al and Si make it challenging to distinguish between them at ever decreasing length\mbox{-}scales. We used the ThermoFisher Helios 650 Nanolab SEM/FIB (ThermoFisher Scientific, Hillsboro, OR) for milling and imaging. The ion beam operated at an energy of 30 keV and the electron beam at an energy of 5 keV. We selected a region at the center of the laser surface remelted (LSR) zone of the same Al\mbox{-}Si sample described in Ref.~\cite{lien_microstructure_2020}, see \textbf{Fig.~\ref{F1_FIB}(a)}. In total, we collected 512 images with 6.5~nm pixel size at 5~nm intervals along the milling direction (corresponding to the laser direction). See \textbf{Fig.~\ref{F1_FIB}(b)} for a representative FIB cross\mbox{-}sectional image and \textbf{Fig.~\ref{F1_FIB}(c)} for an orthogonal view. 

Following data acquisition, we removed ion milling ``curtain" artifacts in each cross\mbox{-}sectional SEM image using a compressed sensing stripe removal algorithm \cite{schwartz_hovden_stripe_removal_2019}. \textbf{Fig.~\ref{F1_FIB}(b)} is a representative image after stripe removal, showing Si fibers with circular cross\mbox{-}section. Then, we stretched all images vertically to account for the 52\textdegree~sample tilt during data acquisition and registered the images using phase correlation in Matlab \cite{MATLAB}, considering only rigid translations. After processing the data, we utilized Zeiss Zen Blue 3.1 with the Intellesis deep learning module \cite{andrew_usage_2017,volkenandt_machine_2018} to perform semantic segmentation, a pixel\mbox{-}based machine learning method, to identify the solid phases in each SEM cross\mbox{-}section. For this purpose, we supplied five  manually segmented images as training data. After training, the model was capable of distinguishing Si from Al. \textbf{Fig.~\ref{F1_FIB}d} shows a reconstructed subvolume, depicting the complex and bicontinuous fibrous Si (see  \textbf{Video.~\ref{V_S1}} for a visualization of a larger volume).
%The [morphological descriptors] of the Si coral structure are quantified from the region of interest shown in Fig 1 and compared to XXX  in Table 1. [Cite Image processing toolkit]. 

\begin{figure}
    \centering
    \includegraphics[scale=.5]{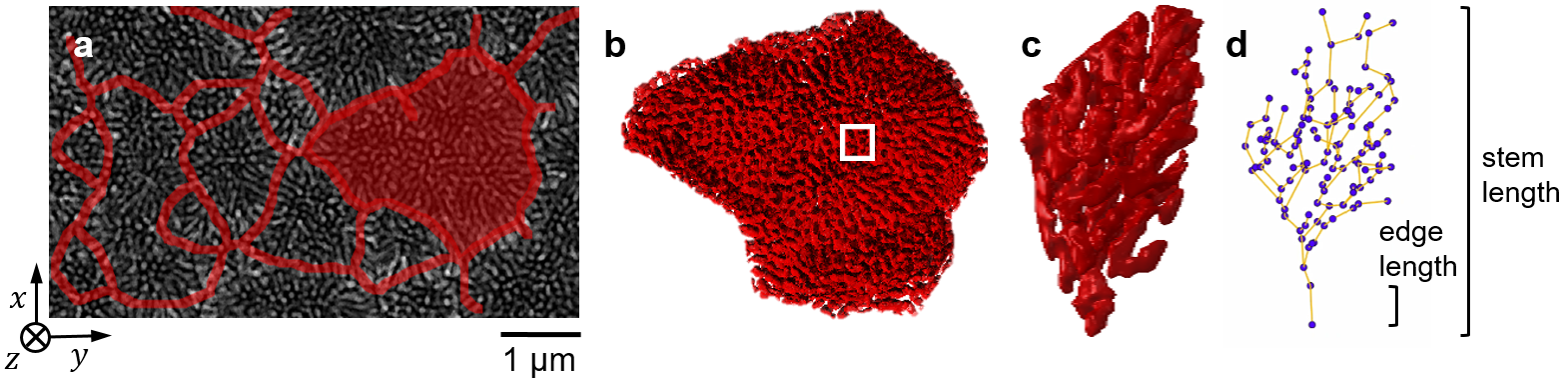}
    \caption{\textit{Hierarchical eutectic microstructure:} (a) a 2D slice of the FIB tomography dataset with lines (in red) that correspond to the boundaries of the colonies. We use a machine learning procedure to identify the lines (see text for details). (b) Single colony that we extract from the 3D data, the same as that shaded in (a). (c) Single tree-like structure within (b) that is branching out isotropically in 3D, from bottom\mbox{-}to\mbox{-}top. (d) Skeleton of the tree structure in (c). The tree can be decomposed into nodes (blue) connected by links (red). The \textit{edge~length} is defined as the length of each link between adjacent nodes and the \textit{stem~length} is the distance from the starting ``seed" of a single tree (at bottom) to all the end\mbox{-}points of the branches.}
    \label{F2_colony}
\end{figure}

%\subsection{Tracing boundaries of eutectic colonies}
Due to the interconnectivity of the ultra\mbox{-}fine Si structure, it is impossible and also unreasonable to interrogate the morphology without having a clear starting and end point within the dataset. To make the problem tractable and meaningful, we decide to isolate a single \textit{colony} for further analysis (as opposed to the aggregate network). \textcolor{black}{These structures form as a result of growth front instability when the ratio of the velocity $V$ to the thermal gradient $G$ exceeds a critical value, thus destabilizing the otherwise planar front \cite{weart_eutectic_1958, yue_controlled_1968}. They are common in laser treated materials since the generally high $V/G$ ratio establishes proper conditions for the formation of cells, even in high\mbox{-}purity alloys, see, \textit{e.g.}, \cite{abboud2020developing}.\footnote{\textcolor{black}{While we have not conducted a detailed chemical analysis, we can estimate the concentration of impurities in our alloy using the constitutional supercooling criterion~\cite{hecht2004multiphase}: assuming $V = 30$~mm/s (from Table~\ref{tab:my_label}) and $G = 300$~K/mm (order-of-magnitude, from Ref.~\cite{kayitmazbatir2022effect}), and taking the thermophysical parameters from Ref.~\cite{liao2014effects}, we would expect  $\sim$10 ppm Sr is enough to destabilize the Al-Si eutectic growth front and generate cells \textit{via} constitutional supercooling. That said, this low concentration is well below that needed for \textit{impurity modification}, which gives rise to refinement and twinning (for an Al20wt\%Si alloy, the critical concentration of Sr for modification is greater than 100 ppm, according to thermodynamic calculations in Ref.~\cite{eiken2015impact}).}} The cells or colonies have} a coarser eutectic microstructure at their boundaries compared to their interior \cite{weart_eutectic_1958}. %In addition, the directionality of the microstructure in the colony is determined by the fact that the eutectic must accommodate the curvature of the solid-liquid interfaces in solidification~\cite{weart_eutectic_1958,hunt1966binary}. 
Unfortunately, however, attempts at manually outlining the colonies were tedious and subject to bias. To overcome this challenge, and inspired by Ref.~\cite{friess2022microstructure}, we quantify several microstructural descriptors of each connected component of Si (\textit{e.g.}, its volume, surface area, and aspect ratio)  \textit{via} Matlab \cite{MATLAB}. Then, we determine the contribution of each descriptor (after normalization) to the overall variance, \textit{via} principal component analysis. This enables us to identify those features that distinguish the Si fibers at the boundary from all  others and ultimately to trace the colony boundaries in 3D, see supplementary information, \textbf{Fig.~\ref{S1_cost}}. A selected colony is highlighted in red in \textbf{Fig.~\ref{F2_colony}(a)} and visualized in 3D in \textbf{Fig.~\ref{F2_colony}(b)}. %generate a linear cost function using the weights  in the first principal component, see  supplementary information. A higher value of the cost function indicates that the Si fiber is located along the periphery of the colony (\textbf{Fig.~\ref{S1_cost}}). By repeating this procedure for all chunks of data, we can trace the colony boundaries in 3D. In \textbf{Fig.~\ref{F2_colony}(a)}, a selected colony is highlighted in red on top of a 2D map of the cost function. The same colony is visualized in 3D in \textbf{Fig.~\ref{F2_colony}(b)}.   

%\subsection{Extracting tree-like structures}
Subsequently, we skeletonize the single colony containing ultra\mbox{-}fine Si fibers \textit{via} medial axis thinning \cite{kollmannsberger_2017_skeletonization}, in order to quantify the branching behavior. The decomposed network structure can be represented by a list of \textit{nodes} (with a defined spatial location) and \textit{links} (spanning two nodes). Despite this reduced representation, we obtain a ``forest" of nodes and links for the highly interconnected Si fibers within the colony. To make our task more manageable, we subdivide the Si fibers inside the colony into individual ``trees" following two empirical rules (see \textbf{Fig.~\ref{S2}} and \textbf{Video.~\ref{V_S1}}). Each tree has characteristic \textit{edge lengths} and a \textit{stem length} (\textbf{Fig.~\ref{F2_colony}(c,d)}), defined in the same way as in Ref.~\cite{gaiselmann_competitive_2013}. 

%Additionally, we determined the angle of branching with regard to the growth direction is calculated to verify the extent to which the direction of growth is in line with the thermal gradient. This also serves as a back-up for the physical meaning behind our rules of separation.

%It's necessary to break connections that don't necessarily reveals the branching morphology of Si rods such as those accidentally formed between touching branches from neighbouring trees.
%and select out trees expanding all the way through as representatives for the tree morphology. We then extract a list of nodes and links for each single tree and obtain statistics including

%\section{Results and discussion} \label{sec:resultsanddiscussion}

We are now positioned to quantitatively compare the quench\mbox{-}modified eutectic  against the impurity\mbox{-}modified eutectic (from Ref.~\cite{gaiselmann_competitive_2013}), with respect to length\mbox{-}scale and morphology. %The fibrous eutectic Si produced \textit{via} both impurity\mbox{-} and quench\mbox{-}modification bear a resemblance at different length scales. We are now positioned to test this idea more rigorously, using the wealth of 3D data with nano\mbox{-}scale resolution on the quench\mbox{-}modified  microstructure (see also \textbf{Video~\ref{V_S1}} for an animation of the growth process, assuming the eutectic grows along the specimen $z$ direction). Additionally, with our novel data processing procedure (\textit{vide~supra}), we can distinguish and extract detailed morphological information from a single eutectic colony. In this way, we can quantitatively compare our work against that of Ref.~\cite{gaiselmann_competitive_2013}, with respect to length\mbox{-}scale and morphology.

%mechanism for impurity modification is , and it has long been assumed that both modification processes leads to a similar microstructure. This leads us to beyond the flake-to-fiber transition of Si eutectics.

%we are lacking insight into rapid quenching due to the limitation of processing techniques. Scientists have long been assuming that both modification processes lead to a similar morphology of Si fibers beyond the flake-to-fiber transition of Si eutectic \cite{hosch_effect_2010,hellawell_growth_1970}. To test the validation of the prediction, we quantitatively compare two solidification samples obtained from both quench- and impurity-modification via length scale and morphology. The morphological data representing the microstructure of impurity-modified Si eutectic analyzed in the following section are obtained from Gaiselmann's published paper from his paper, we extracted the curves in the plots that provides us with the branching statistics for our comparison. \cite{gaiselmann_competitive_2013}. 

%\subsection{Length\mbox{-}scale} \label{sec:lengthscale}

\begin{figure}[h]
    \centering
    \includegraphics[scale=.6]{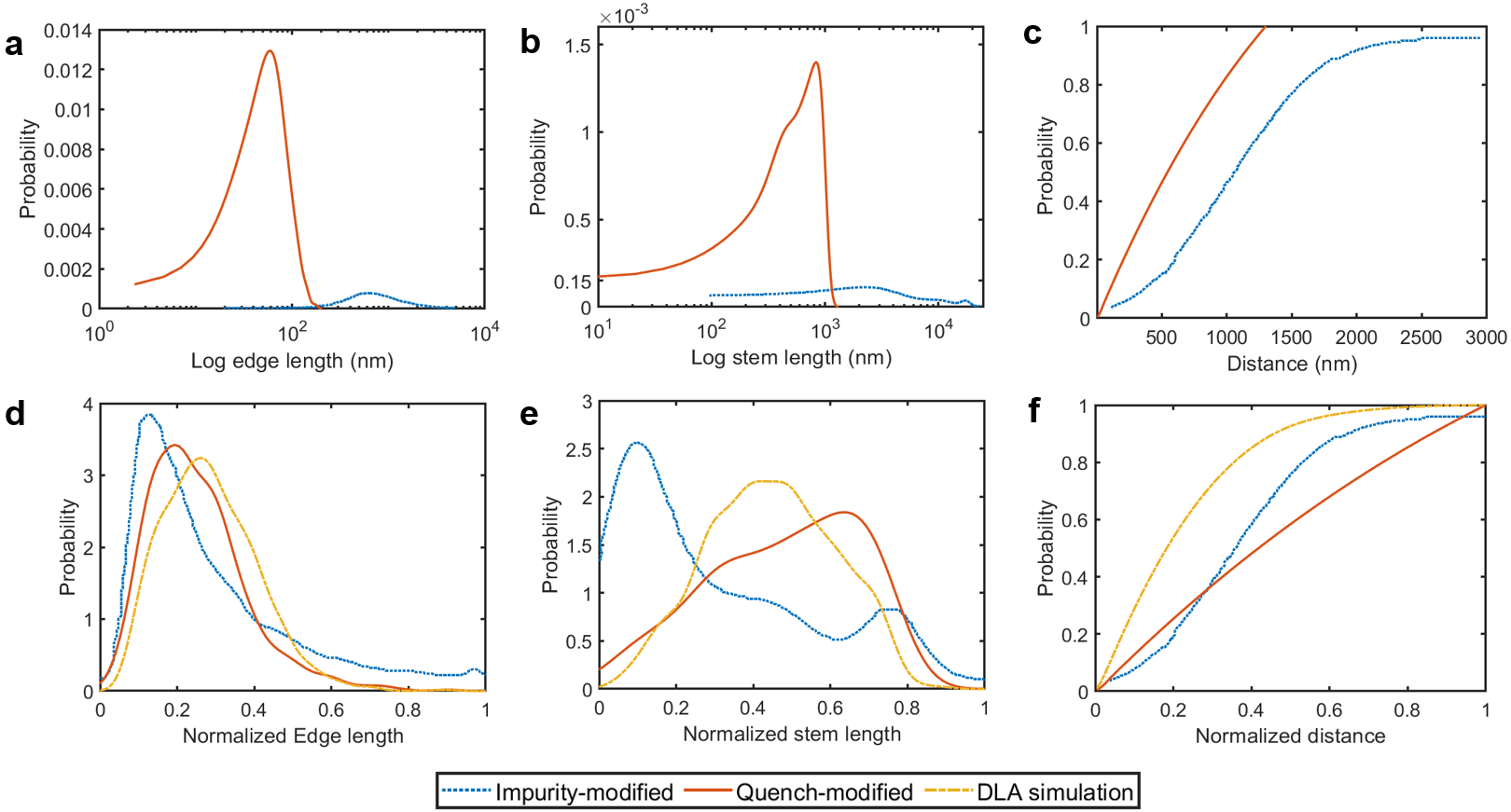}
    \caption{\textit{Quantitative comparison of the tree\mbox{-}like structures} according to three metrics: (a,d) edge lengths, (b,e) stem lengths, and (c,f) contact distributions. Data presented on histograms come from quenched- (red) and impurity-modified (blue) structures. (a-b) are plotted on  a log-linear scale due to the significant difference in length scale between the two. Top row shows length\mbox{-}scales in physical coordinates while bottom row is normalized to control for differences in length\mbox{-}scale.  On the bottom plots we give also distributions obtained from DLA simulation (yellow) using a sticking coefficient of 0.4, see text for details.}    
    \label{3}
\end{figure}

%We measure the edge and stem lengths from tree structures in the eutectic colonies (defined in \textbf{Fig.~\ref{F2_colony}(d)}).
\textbf{Fig.~\ref{3}(a,b)} shows  distributions of edge and stem lengths, respectively. We also quantify \textit{via} cross\mbox{-}correlation the contact distribution function $H(r)$ in \textbf{Fig.~\ref{3}(c)}, defined as the probability that the minimum distance from a random position in the Al matrix to the Si tree structure is less than $r$ for $r>0$~\cite{gaiselmann_competitive_2013}. Superimposed are the same descriptors from Ref.~\cite{gaiselmann_competitive_2013} on the impurity\mbox{-}modified eutectic. Unsurprisingly, both edge and stem lengths of the impurity\mbox{-}modified structure are on the order of micrometers, while the quenched structure shows finer features down to the nanometer length\mbox{-}scale. In fact, we find an average fiber radius of 22.5~nm and an average fiber spacing $\lambda$ of 91.0~nm for the quenched structure using the auto\mbox{-}correlation function (see \textbf{Fig.~\ref{S4}}). This is near the finest possible spacing of 39.4 nm predicted by Ref.~\cite{wang_limit_2011} as the limit of steady\mbox{-}state eutectic growth\textcolor{black}{, see \textbf{Table~\ref{tab:my_label}}}. Finally, we report a surface density of 58.0307 $\mu$m$^{-1}$, over 100 times larger than that of the impurity\mbox{-}modified sample. \textbf{Table~\ref{table}} summarizes the data from both eutectic structures.

In general, the nanoscale features of Si fibers in the LSR meltpool can be attributed to the deep undercooling and high growth velocity. Simulations \cite{bollada_faceted_2018} and experiments \cite{liu_morphologies_nodate,panofen_solidification_2007,yang_dependence_2011} have shown that a larger undercooling results in ``facet breaking" which, for eutectics, should coincide with a transition from flake\mbox{-}like to fibrous Si. Adopting the  undercooling\mbox{-}spacing\mbox{-}velocity relationships modified for rapid solidification \cite{pierantoni_coupled_1992}, we estimate a growth velocity of 30.2~mm~s$^{-1}$ (which, fortuitously, corresponds to the laser scanning speed) and an undercooling of 72.7~K. In comparison, Hosch and coworkers~\cite{hosch_analysis_2009, hosch_solidification_2010}~reported that the flake\mbox{-}to\mbox{-}fiber transition of Si occurs in two stages with the first being the in-plane break\mbox{-}down of Si plates and the second being the out\mbox{-}of\mbox{-}plane Si branch growth. This conclusion was drawn from deep etching experiments; as such, it is difficult to ascertain whether fiber growth above the second transition is truly 3D in nature or is confined within the Si plates, as the authors themselves acknowledge~\cite{hosch_analysis_2009}. Nevertheless, our calculated velocity is far above that of the second transition (see \textbf{Table~\ref{tab:my_label}}), which may explain the fibrous structure in \textbf{Fig.~\ref{F2_colony}c\mbox{-}d}. %from our sample. This is beyond the transition line between between a fully fibrous Si eutectic and fibrous eutectics with Al dendrites for a 16 wt\% Si sample \cite{pierantoni_coupled_1992}.  

\begin{table}[]
    \centering   
    
    \begin{tabular}{lllllll}
\hline & $f_{\mathrm{Si}}$ & $\mathrm{wt} \% \mathrm{Si}$ & $\lambda(\mu \mathrm{m})$ & $s_{S}\left(\mu m^{-1}\right)$ & $r_{e q}(\mu \mathrm{m})$ & Ref. \\
\hline Quench-modified & $0.1805$ & $\textcolor{black}{20}$ & $0.091$ & $58.0307$ & $0.0225$ & This work \\
Impurity-modified & $0.14$ & $\textcolor{black}{7}$ & $\sim 2$ \textsuperscript{*} & $0.527$ & $0.55$ & \cite{gaiselmann_competitive_2013} \\
DLA & $0.3512$ \textsuperscript{**} & $\textcolor{black}{36}$ & $\mathrm{n} / \mathrm{a}$ & $\mathrm{n} / \mathrm{a}$ & $\mathrm{n} / \mathrm{a}$ & This work \\
Eutectic point (nominal) & $0.143$ & $\textcolor{black}{12.6}$ & $\mathrm{n} / \mathrm{a}$ & $\mathrm{n} / \mathrm{a}$ & $\mathrm{n} / \mathrm{a}$ & \cite{Massalski1986BinaryAP} \\
\hline
\end{tabular}

\caption{\textit{Basic descriptors of eutectic microstructure produced by different routes}: $f_{Si}$ is the volume fraction of Si phase \textcolor{black}{(directly measured from the segmented volume, in our work)}, wt\% is weight percent, $\lambda$ is the fiber spacing \textcolor{black}{(defined as the distance between neighboring stems of Si)}, $s_S$ is the specific surface density of Si phase \textcolor{black}{(the surface area of Si phase divided by its volume)}, and $r_{eq}$ is the radius of Si fibers. \textcolor{black}{Based on past reports~\cite{grugel1987growth}, $\lambda$ does not vary with composition, holding all else constant.}\\ \textsuperscript{*}: Value is calculated from the contact distribution function $H(r)$ provided in Fig.~12 of Ref.~\cite{gaiselmann_competitive_2013}.\\ \textsuperscript{**}: The relatively large Si phase fraction obtained from DLA simulation is caused by dilating the tree structure to match the 22.5~nm fiber radius from the quenched sample.}

    \label{table}
    
\end{table}

\begin{table}[]
    \centering 
    
    \begin{tabular}{llllll}
\hline & $\lambda$ $(\mathrm{nm})$ & V $(\mathrm{mm} / \mathrm{s})$ & $\Delta$T (K) &  Ref \\
\hline Rapid quenching & $91.0$ & 30.2 & $72.7$ &  This work\\
Coupled zone boundary, for 16~wt\%Si & $65.3$ & $46.2$ & $120.6$ &  \cite{pierantoni_coupled_1992}\\
 Limit of fiber spacing & $39.4$ & 133 & $240.2$ &  \cite{wang_limit_2011}\\
Upper limit of flake-to-fiber transition & $11-15 \times 10^{3}$ \textsuperscript{***} & $0.5-0.95$ & $\sim 10 ^*$ &  \cite{hosch_analysis_2009}\\
\hline
\end{tabular}
   \caption{\textit{Solidification parameters} for this study (top row) compared with other critical values from the literature (below). \textsuperscript{***}:~Value is estimated from the critical velocity obtained in Ref.~\cite{hosch_analysis_2009}.}
    \label{tab:my_label}
\end{table}

%\subsection{Morphology}
Aside from the obvious differences in length\mbox{-}scale, it remains to be determined whether the two processing routes lead to the same morphology of the Si phase. To answer this question, we plot the three histograms again, this time normalizing the length\mbox{-}scales to a range between zero and one (see \textbf{Fig.~\ref{3}(d-f)}). The difference in the two distributions remains even \textit{after} controlling for length\mbox{-}scale, indicating two fundamentally different morphologies with different origins. We conduct a two\mbox{-}sample Kolmogorov–Smirnov (K-S) test to quantify the statistical distance between the two distributions~\cite{massey1951kolmogorov, miller1956table, marsaglia2003evaluating} and ultimately reject the null hypothesis that the two curves are drawn from the same distribution, at a 5\% significance level. It follows that the two microstructures are \textit{not} self\mbox{-}similar.

\begin{figure}[ht]
    \centering
    \includegraphics[scale=.399]{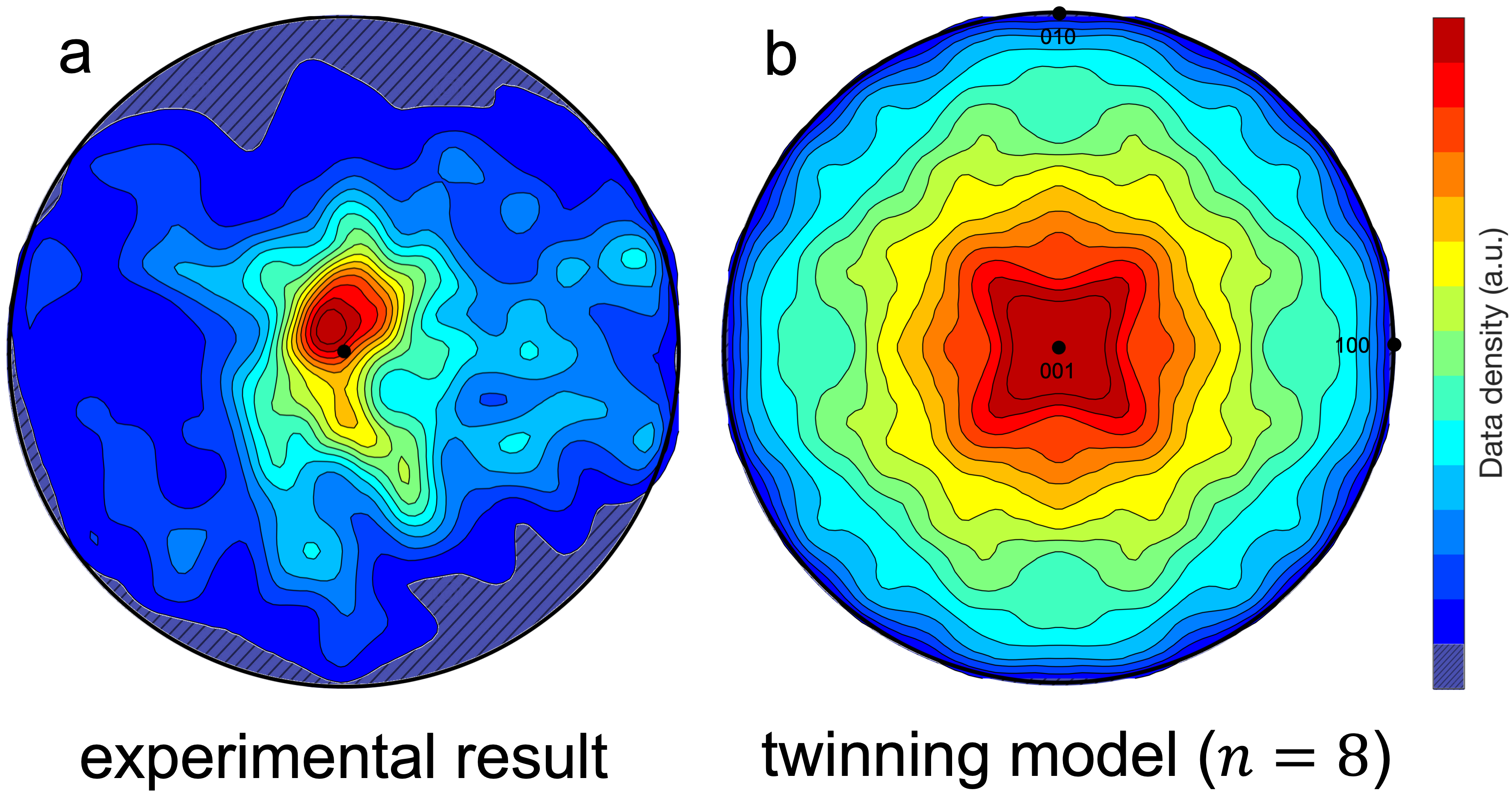}
    \caption{\textit{Distributions of Si fiber growth directions}: comparison between (a) experimental result on ultra-fast quenching and (b) theoretical prediction of the branched morphology caused by multiple twinning on \{111\} planes. Zone axis (black dot) in stereographic projections is specimen $z$ (cf.~\textbf{Fig.~\ref{F1_FIB}}) and crystallographic $\langle100\rangle$ direction, respectively. Cross-hatched regions include an absence of data.}
    \label{F4}
\end{figure}

Instead, we posit that rapid cooling in LSR not only produces a much finer eutectic microstructure, but also a highly interconnected and isotropic branching behavior. To test this idea more rigorously, we compared stereographic projections of the Si fiber edge directions in the quenched structure against a theoretical prediction from multiple twinning, see \textbf{Figs.~\ref{F4}}. The latter is determined by calculating the total number of unique crystal orientations $N(n)$ that are possible after $n$\textsuperscript{th}-order twinning on \{111\} planes, which follows a geometric series \cite{hellawell_growth_1970, shahani2016interfacial}: $N(n)=4 \sum_{t=1}^n 3^{t-1}+1$. For example, after $n=8$ repeated twinning events, there are a total of $N(8)=13121$ unique crystal orientations (and hence growth directions) that cover the $\langle100\rangle$ stereographic projection, although \textit{not} uniformly. That is, \textbf{Fig.~\ref{F4}(b)} shows a preferred four-fold symmetry, which reflects the cubic symmetry of the Si crystal. In contrast, the fiber growth directions are non\mbox{-}directional or isotropic in \textbf{Fig.~\ref{F4}(a)}.  

We rationalize the above observations by noting that more growth sites are operative at high undercooling that are otherwise inaccessible at lower undercooling \cite{napolitano_faceted_2004}.  As more and more sites become active, the kinetic factor of Si in the revised Wilson-Frenkel law~\cite{JACKSON196753} will approach that of typical metals. That is, provided that the undercooling is above the threshold for the flake\mbox{-}to\mbox{-}fiber transition (as it is in our case), growth on multiple sites will naturally lend itself to a more isotropic morphology (namely, branching in any and all directions \textit{irrespective} of the underlying crystallography of Si). An alternative explanation is that the kinetic undercooling of the Si phase increases with growth rate, such that the lead distance of the Si phase at the growth front would be gradually reduced.  Eventually the Al and Si phases would be fully coupled at the growth front, such that the shape of the Si phase would no longer be governed by faceted growth~\cite{major1989effect,mohagheghi2020decoupled}. 

In contrast, theories for impurity\mbox{-}induced modification argue that the impurity atoms (such as Na or Sr) are rejected by the solidification front; accumulate at the eutectic\mbox{-}liquid interface; interfere with the layer growth of Si (once the eutectic Si has grown far enough to collect sufficient impurity atoms); and increase the number of twin defects in Si, thus enabling more frequent branching \cite{lu1987mechanism, major1989effect,moniri2019chemical}. This cycle of crystallographic branching occurs with a characteristic frequency that is given by the edge length in \textbf{Fig.~\ref{3}(a,d)}. That said, branching is still constrained by the favored growth directions of Si and the locations of the twin plane grooves (provided the alloy is solidified under low growth rates)  \cite{wagner1960growth,hamilton1960propagation,day1968microstructure}. 

%reason(references)

%\subsection{Diffusion-limited aggregation}
%Broader context: existing structure also showing isotropic growth (random fractal)

We put the above result in a broader context by comparing, for the first time, the morphology of Si to another, prototypical branched structure produced by diffusional aggregation: a random fractal~\cite{Witten_1981_DLA, bunde2012fractals}. We follow standard algorithms for DLA \cite{PhilipThomas_2012, Witten_1981_DLA} to simulate the latter \textcolor{black}{and test for its similarity to Si morphology under quenching}. We restrict the movement of random walkers along the specimen $z$ direction to match the growth direction of the eutectic in experiment. We use the K-S test again to determine the sticking coefficient (the probability that a walker will stick to the aggregate) that resembles our quenched tree structure most closely, see \textbf{Fig.~\ref{S3}}. Then, as before, we obtain statistics on edge length, stem length and contact distribution (yellow curves in \textbf{Fig.~\ref{3}(d-f)}). 

%Data on edge length and stem length from DLA resembles more that of the quenched-modified structure on visual inspection. Indeed, 
According to the K-S test, the edge lengths of the random fractal and the quench\mbox{-}modified structure come from the same distribution at a 5\% significance level. We reject the null hypothesis on stem length, however, although we suggest that this discrepancy is because the stem length is sensitive to the volume of the colonies from which the distributions are obtained. Likewise, the contact distributions of the two structures show a clear difference. This is due to the physics of directional eutectic solidification.\footnote{See Ref.~\cite{das2002simulation} for the multi\mbox{-}particle extension of DLA for the simulation of solidification structures.} That is, with the high thermal gradient of LSR, the Si fibers are aligned with the macroscopic growth direction (6.4\% of edges are within 15$^{\circ}$ of $z$ \textit{vs.}~3.4\% for a random distribution, see also \textbf{Fig.~\ref{F4}(a)}). This would imply that the contact distribution $H(r)$ increases linearly with normalized distance $r$. Taken altogether, our analysis of the descriptors indicates that the ultrafine Si in LSR is in its \textit{own} morphological class, distinct from impurity\mbox{-}modified and random fractal structures.\footnote{Pursuing also a comparison of the three histograms for the impurity-modified and random fractal structures, we find a significant difference at the 5\% confidence level according to the K-S statistic.} % This new insight can inform future design of eutectic components produced from laser\mbox{-}based manufacturing processes.

%\section{Conclusion}
In this paper, we characterize in 3D the morphology of an ultra\mbox{-}fine Al-Si eutectic processed \textit{via} laser rapid solidification. We map the fibrous Si microstructure to a network of nodes and links to quantify its branching behavior. This representation brings to light that the quench\mbox{-}modified eutectic not only has a finer spacing (on the nanoscale) compared to its impurity\mbox{-}modified counterpart, but perhaps more surprisingly, that the morphologies of the two eutectics are not necessarily the same. For example, the branching behavior of the quench\mbox{-}modified eutectic is remarkably isotropic (somewhat akin to a random fractal produced \textit{via} DLA). This would indicate the absence of Si facets exposed to the liquid, the result of either kinetic roughening or coupled growth at deep undercooling.  \textcolor{black}{Further insight may be gained from examining the Si morphology of impurity\mbox{-}doped samples under LSR, for which experiments are well underway.} Broadly, our quantitative analysis can be used to generate synthetic eutectic microstructures that reflect the inherent complexity of real ones.

%We suggest this is to be a result of more activated growth directions under high undercooling compared with those allowed by impurity addition. We also did DLA simulations to verify the higher degree of randomness observed in quenched structure and we did observe a 
\section*{Acknowledgements}
We  gratefully  acknowledge  financial  support  from  the  National  Science  Foundation  (NSF)  CAREER  program  under  award  number~1847855. \textcolor{black}{Amit Misra and Huai\mbox{-}Hsun Lien thank the support from DOE, Office of Science, Office of Basic Energy Sciences, grant number~DE\mbox{-}SC0016808 for the laser processing work. We acknowledge assistance from Metin Kayitmazbatir and late Prof. Jyoti Mazumder in laser scanning experiments.} We thank the University of Michigan College of Engineering for financial support and the Michigan Center for Materials Characterization for use of the instruments and staff assistance.

%directional growth & 3D isotropy

%% If you have bibdatabase file and want bibtex to generate the
%% bibitems, please use
%%
 %\bibliographystyle{elsarticle-num} 
 \bibliographystyle{elsarticle-num-names}
 \bibliography{AlSi}

%% else use the following coding to input the bibitems directly in the
%% TeX file.

% \begin{thebibliography}{00}

% %% \bibitem{label}
% %% Text of bibliographic item

% \bibitem{}

% \end{thebibliography}

%% The Appendices part is started with the command \appendix;
%% appendix sections are then done as normal sections
\newpage
% \appendix

\section*{Supplementary material}
\label{supplement}

\renewcommand{\thefigure}{S\arabic{figure}}
\setcounter{figure}{0}
 
\renewcommand{\theVideo}{V\arabic{Video}}
\setcounter{Video}{0}

\begin{Video}
% \captionsetup{font=normalsize}
\caption{\textit{Visualization of the 3D fibrous nano-scale eutectic.} The first\mbox{-}half of the video animates the growth process from different perspectives. The second\mbox{-}half shows a single tree-like structure of Si, and its associated nodes and links, used to quantify the branching behavior (shown also in \textbf{Fig.~\ref{F2_colony}c-d}).}   \label{Animation} 
\label{V_S1}  
\end{Video}

\begin{figure}[h]
    \centering
    \includegraphics[scale=.5]{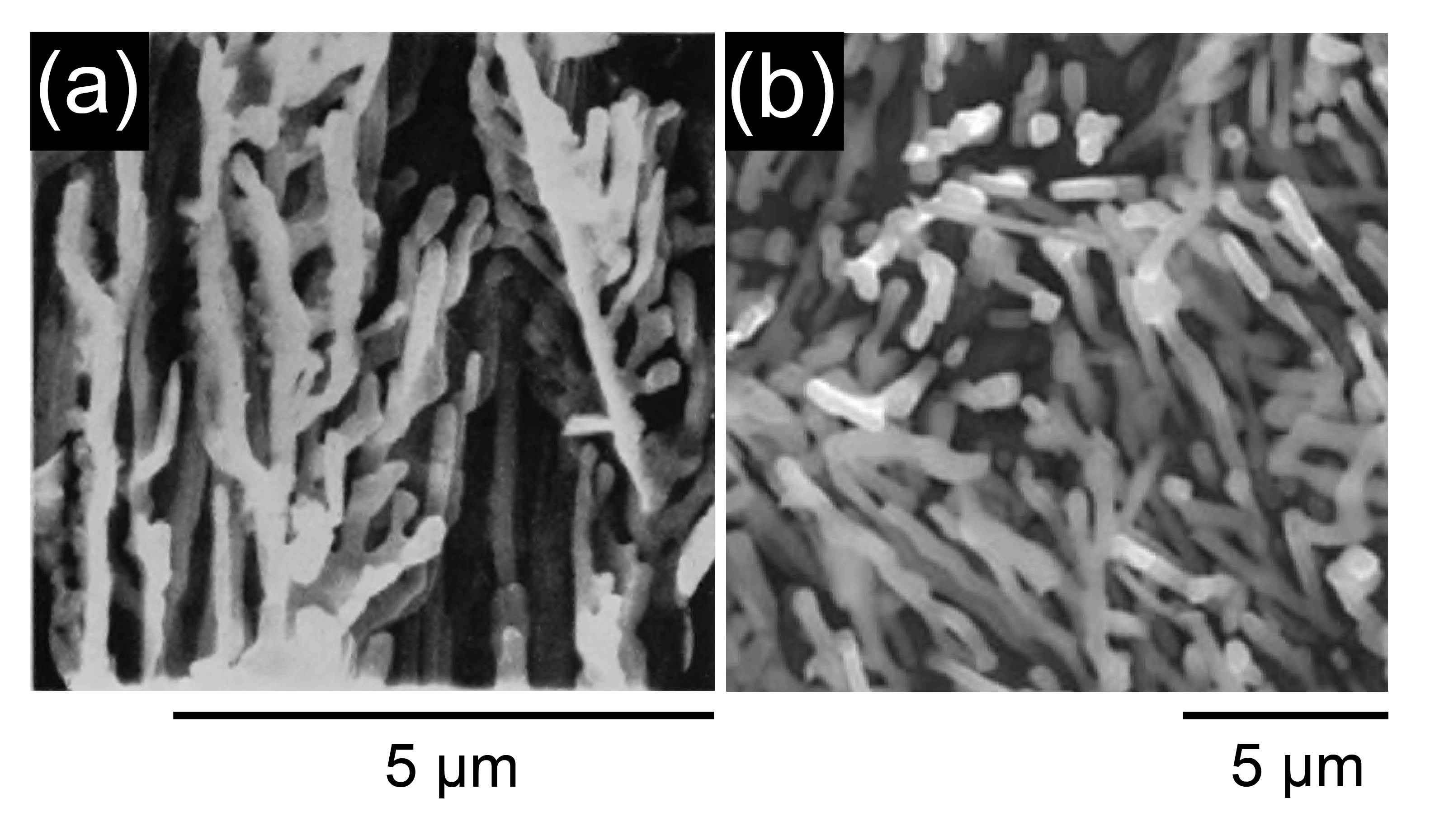}
    \caption{\textit{Side-by-side visual comparison} of the impurity modified  \cite{day1968microstructure} and quench modified fibrous structure of Si phase \cite{hosch_analysis_2009}. The Al phase has been chemically removed in both images.  Adapted with permission.}
    \label{S0_compare}
\end{figure}

\clearpage
\subsection*{Procedure for tracing colony boundaries}
% the three sections are still kinda separate taking about the figures
We decompose the microstructure into colonies by identifying the Si fibers that are located along the colony boundaries. According to the arguments given in Ref.~\cite{weart_eutectic_1958}, fibers along the boundaries tend to be coarser (and hence have a larger eutectic spacing). With this in mind, we selected eight descriptors to analyze the ensemble of fibers in the microstructure: ($i$) volume, ($ii$) surface area, ($iii$) ratio of the largest and smallest principal axis length in 3D,  ($iv$) solidity, as well as ($v-viii$) the maximum, minimum, total, and mean distances between fibers and all of their nearest neighbors. The nearest neighbours are defined as those fibers closest to the target fiber in all directions. To identify the descriptors that help to distinguish between fibers at the boundary \textit{vs.}~those at the interior, we perform a principal component analysis. The first principal component (PC1) is a combination of volume $V$, surface area $S$, and aspect ratio $\phi$; it contributes more than 90\% of the variance. Accordingly, we define a cost function to be a linear function of these three parameters, where the coefficients are taken as the weights $w$ in PC1, \textit{i.e.}, $\mbox{cost}=w_V V + w_S S + w_\phi \phi$. A higher cost value indicates that the Si fiber is located along the periphery of the colony.

\begin{figure}[h]
    \centering
    \includegraphics[scale=.65]{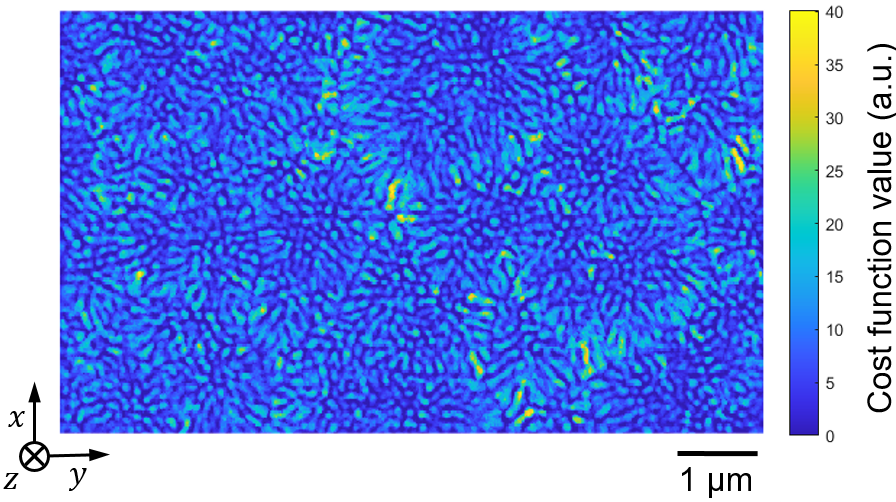}
    \caption{\textit{2D slice from the 3D dataset,} wherein the Si fibers are colored according to value of cost function. }
    \label{S1_cost}
\end{figure}

We can color each Si fiber according to its cost in the cross\mbox{-}sectional images ($x$-$y$ plane). The outline of the colonies can then be reasonably identified from a projection along the image depth ($z$) for ``chunks" of 30 cross\mbox{-}sectional images, see one such projection in \textbf{Fig.~\ref{S1_cost}}.

%To further enhance the difference between objects that are along the boundaries from those at the center of a colony, we projected the result of cost function from 3D volumes into 2D plane by adding the value of cost function of 31 slices altogether and project the sume onto the middle plane (the 16th in the stack). The number of 31 more slices is the minimum necessary to obtain a reasonable outline of colonies. \textbf{Fig.~\ref{S1_cost}} shows one such slice where the Si rods are colored according to the cost function; we obtained this 2D map from a stack of 31 slices with the shown one in the middle. It can be seen that the highlighted yellow rods decorate the boundaries of the colonies. By processing all the slices? in this way, we can trace the boundaries in 3D.

\clearpage
\subsection*{Procedure for identifying tree\mbox{-}like structures}

\begin{figure}[h]
    \centering
    \includegraphics[scale=.5]{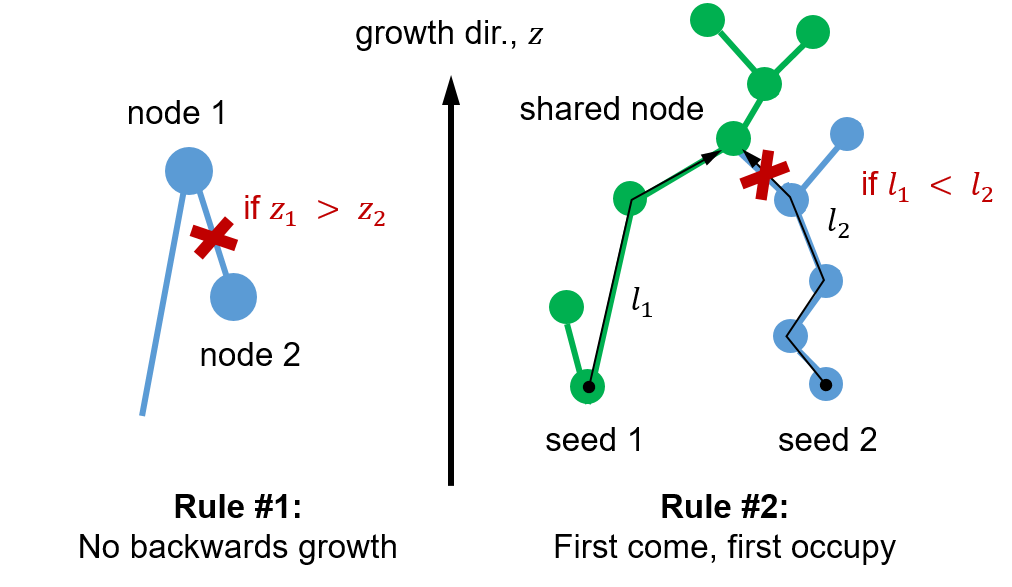}
    \caption{\textit{Schematic of the two empirical rules} that were enforced in order to separate the tree structures in the quenched sample. Here, $z_i$ is the $z$ coordinate of node $i$ (along the growth direction) and $l_i$ is the length of the path (in black) between seed $i$ and the shared node. See text for details.}
    \label{S2}
\end{figure}

To subdivide the colonies into ``tree"-like structures of connected fibers, we developed two empirical rules that are consistent with the physics of directional eutectic solidification. These rules enabled us to  break links that connected two trees (due to merging of branched fibers in solidification~\cite{steinmetz2018graph}) and that ultimately rendered a single connected component (instead of two separate ones). \textbf{Fig~\ref{S2}} provides a schematic illustration of both rules:
\begin{enumerate}
    \item \textit{Rule \#1: No backwards growth.} Here, we break links that point backwards (along $-z$), \textit{i.e.}, away from the macroscopic growth direction ($+z$). Physically, it is not possible for a fiber to grow into a region that has already solidified. 
    \item \textit{Rule \#2: First come, first occupy.} Here, we consider nodes that are shared between two or more trees. We assign them to the tree that arrives at that node earlier than the others, \textit{i.e.}, the tree that `wins' is likely the one that has a shorter path from the starting seed to the shared node.  The path length is calculated as the total length of the previous links from the starting seed up to the shared node. This rule follows from the fact that Si fibers can impinge upon fibers that were solidified beforehand. 
\end{enumerate}

While these rules can also impact the final tree morphology of Si phase, we try to minimize such an effect by selecting out those trees making their way through the whole colony as our target structures when gathering the branching statistics.

\clearpage
\subsection*{Basic descriptors of microstructure}

\begin{figure}[h]
    \centering
    \includegraphics[scale=.6]{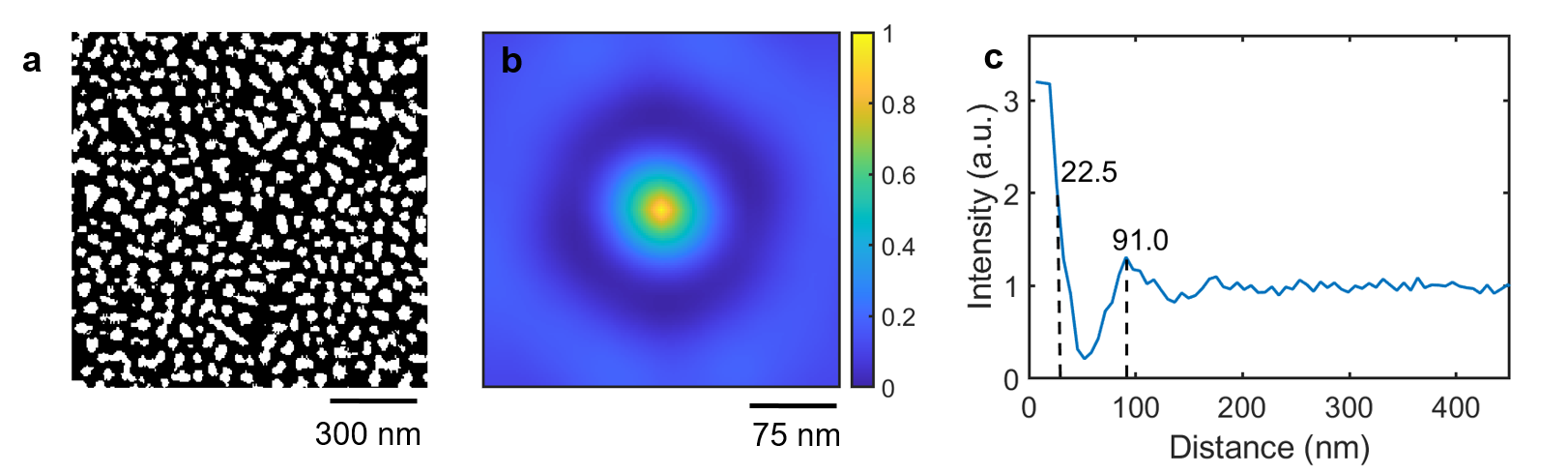}
    \caption{\textit{Determination of basic statistics by auto-correlation function}. (a) Section of one representative 2D slice from the 3D dataset. (b) Density plot of the auto-correlation (a.u.). (c) Radial distribution function, where the origin corresponds to the center of a Si fiber. Accordingly, the average fiber radius and average spacing in this example is 22.5~nm and 91.0~nm, respectively.}
    \label{S4}
\end{figure}

We obtain the Si fiber diameter and spacing by calculating the 2D auto-correlation of Si phase. \textbf{Fig.~\ref{S4}(a)} shows a section of one 2D slice that we use to demonstrate our procedure, taken from the center of a colony. \textbf{Fig.~\ref{S4}(b)} shows the auto-correlation of \textbf{Fig.~\ref{S4}(a)}. The size of the central peak (in yellow, a.u.) indicates the average size of the Si fibers. The distance between the central peak and the first-order ring indicates the spacing between the Si fibers. Note that it is invariant with direction, indicating that the eutectic microstructure is spatially isotropic. \textbf{Fig.~\ref{S4}(c)} shows the radial distribution function, computed from \textbf{Fig.~\ref{S4}(b)} with respect to the distance from the center of a Si fiber. The width of the primary peak gives the fiber diameter as 22.5~nm, and the distance between primary and secondary peaks gives the average spacing as 91.0 nm. 

\clearpage
\subsection*{DLA simulation}

\begin{figure}[h]
    \centering
    \includegraphics[scale=.6]{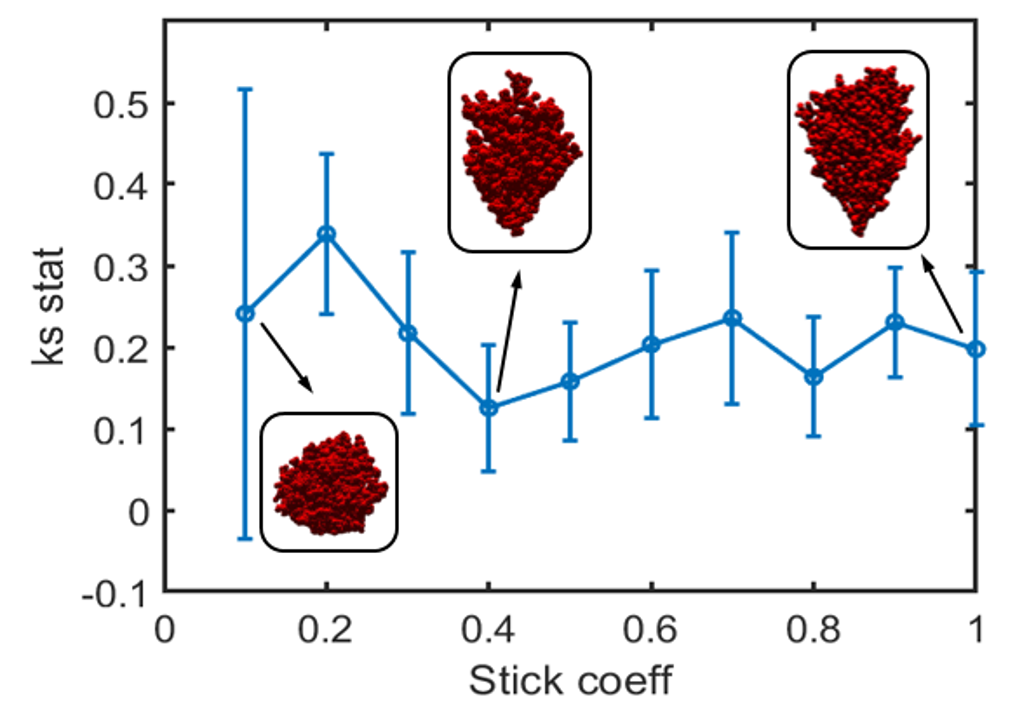}
    \caption{\textit{Similarity measures between the quenched microstructure} vs.~\textit{DLA structures} that were simulated with varying stick coefficients. The  coefficient of 0.4 corresponds most closely to the quenched structure visually and quantitatively by the K-S test, and is the one that we use in our analysis in \textbf{Fig.~\ref{3}(d-f)}. Insets show examples of tree-like structures generated \textit{via} DLA using different sticking coefficients.}
    \label{S3}
\end{figure}

When conducting DLA simulation, the sticking coefficient is one input parameter that influences strongly the structure of the tree clusters formed by random walkers \cite{Witten_1981_DLA, bunde2012fractals}. A sticking coefficient of 1 implies that all walkers will stick to the cluster, and 0 the converse. To determine which coefficient to use, we vary the coefficient from 0.1 to 1 in increments of 0.1 and generate 20 sets of random fractal clusters for each case. We then obtain branching statistics of all DLA tree structures following the same procedure (skeletonization) that we followed for the experimental data.  We again use the K-S statistic to codify the similarity of distributions of edge and stem length between DLA and experimental data on the quench\mbox{-}modified structure. The result is shown in \textbf{Fig.~\ref{S3}} where a lower K-S statistic value indicates a closer agreement in branching statistics. With this data in hand, we select a sticking coefficient of 0.4 to match the experimental data as best as possible.

\end{document}